\documentclass[eqsecnum,showpacs,showkeys]{revtex4}

\usepackage{graphicx}
\usepackage{epsfig}
\usepackage{dcolumn}
\usepackage{amsfonts}
\usepackage{amsmath}
\usepackage{amssymb}
\usepackage{bm}

\begin{document}

\newcommand{\brm}[1]{\bm{{\rm #1}}}

\title{Renormalized field theory of resistor diode percolation}

\author{Olaf Stenull}
\author{Hans-Karl Janssen}
\affiliation{
Institut f\"{u}r Theoretische Physik 
III\\Heinrich-Heine-Universit\"{a}t\\Universit\"{a}tsstra{\ss}e 1\\
40225 D\"{u}sseldorf\\
Germany
}

\date{\today}

\begin{abstract}
We study resistor diode percolation at the transition from the non-percolating to the directed percolating phase. We derive a field theoretic Hamiltonian which describes not only geometric aspects of directed percolation clusters but also their electric transport properties. By employing renormalization group methods we determine the average two-port resistance of critical clusters, which is governed by a resistance exponent $\phi$. We calculate $\phi$ to two-loop order.
\end{abstract}

\maketitle

\section{Introduction}
Percolation~\cite{bunde_havlin_91_etc} is a leading geometric model for connectivity in irregular media. In abstract terms, percolation is the passage of an effect through a medium which is irregularly structured in the sense that the effect can propagate through some regions whereas it cannot pass other areas. Such a medium could be a network of resistors connecting nearest neighboring sites of a $d$-dimensional hyper-cubic lattice and the effect would be a current $I$ injected at site $x$ and extracted at site $x^\prime$. Irregularity in the medium could be realized by randomly removing resistors with a probability $1-p$.

Depending on the occupation probability $p$ the resistors (bonds) are likely to either be isolated or to form clusters. Two sites belong to the same cluster if they are connected by a path of bonds and hence current can flow between them. At low  $p$ two infinitely separated terminal sites $x$ and $x^\prime$ are not connected by such a path and the network behaves as an insulator. For large $p$, on the other hand, many paths between $x$ and $x^\prime$ may exist and the network is a conductor. Therefore, at some probability in between, a threshold $p_c$ must exist where for the first time current can percolate from $x$ to $x^\prime$. Below $p_c$ we have an insulator. Above $p_c$ we have a conductor. The threshold probability is called the percolation threshold, or, since it separates two different phases, the critical probability. 

In this picture the effect can percolate through occupied bonds in all directions. The resulting clusters are typically isotropic in space. Therefore ordinary percolation is referred to as isotropic percolation (IP). The linear extend of the isotropic clusters can be characterized by the correlation length $\xi \sim \left| p - p_c \right|^{-\nu}$, where $\nu$ is the correlation length exponent of the IP universality class.

Directed percolation (DP)~\cite{hinrichsen_2000} is an anisotropic variant of percolation. The bonds function as diodes so that the current can percolate only along a given preferred direction. Typical DP clusters are anisotropic and they are characterized by two different correlation lengths: $\xi_{\parallel}$ (parallel to the preferred direction) and $\xi_\perp$ (perpendicular to it). As one approaches the critical probability, the two correlation lengths diverge with the exponents $\nu_\parallel$ and $\nu_\perp$ of the DP universality class.

DP is perhaps the simplest model resulting in branching self-affine objects. It has many potential applications, including fluid flow through porous media under gravity, hopping conductivity in a strong electric field~\cite{vanLien_shklovskii_81}, crack propagation~\cite{kertez_vicsek_80}, and the propagation of surfaces at depinning transitions~\cite{depinning}. Furthermore, it is related to epidemics with recovery~\cite{grassberger_85} and self-organized critical models~\cite{soc}.

If the preferred direction is interpreted as time, a DP cluster represents the history of a stochastic process. In this interpretation the DP universality class is the generic universality class for phase transitions between an active and an inactive absorbing state.

The transport properties of IP have been studied extensively~\cite{sammlung}. For example, the average two-port resistance $M_R$ of a random resistor network (RRN), i.e., a percolation model in which the bonds are assigned a finite nonzero conductance, is known to obey in the vicinity of the percolation threshold the power law~\cite{harris_fisch_77,dasgupta_harris_lubensky_78}
\begin{eqnarray}
M_R (x,x^\prime ) \sim \left| x - x^\prime \right|^{\phi^{\text{IP}}/\nu}
\end{eqnarray}
with $\phi^{\text{IP}}$ being referred to as the resistance exponent for IP.

Compared to IP relatively little is known about transport in DP. In this paper we study the average resistance in DP by employing the powerful methods of renormalized field theory. We find that this quantity is governed in the vicinity of the critical probability by a resistance exponent $\phi$ analogous to $\phi^{\text{IP}}$. For instance, measured in the preferred, time-like direction between two terminal sites $x = (x_\perp =0, t)$ and $x^\prime = (x_\perp^\prime =0, t^\prime)$ it scales like
\begin{eqnarray}
M_R (x,x^\prime ) \sim \left( t - t^\prime \right)^{\phi/\nu_\parallel} \, .
\end{eqnarray}
We obtain that $\phi$ is given to second order in the departure $\epsilon = 5-d$ from 5 dimensions by
\begin{eqnarray}
\phi  = 1 + \frac{\epsilon}{24} \left\{ 1 + \left[ \frac{151}{288} - \frac{157}{144}\ln \left( \frac{4}{3} \right) \right] \epsilon \right\}  \, .
\end{eqnarray}  

Our field theoretic approach thrives on our real world interpretation of Feynman diagrams which has proven to be successful in studying transport in IP via RRN~\cite{stenull_janssen_oerding_99,janssen_stenull_oerding_99,janssen_stenull_99,stenull_2000,stenull_janssen_2000a,stenull_janssen_2001,stenull_janssen_oerding_2001}. A real-world interpretation in the same spirit applies to DP also. We show that the Feynman diagrams utilized in this paper can be interpreted as being directed resistor networks. On one hand this remedies the complexity of the field theory and makes it more intuitive. On the other hand it simplifies the actual calculations. As shown below, the basic task in calculating the average resistance in DP is to calculate the resistance of Feynman diagrams. The remaining steps are completely analogous to those well known from previous field theoretic approaches~\cite{cardy_sugar_80,janssen_81,janssen_2000} to usual, i.e., purely geometric DP. 

Brief account of our results was given in~\cite{janssen_stenull_directedLetter_2000}. The purpose of this paper is to present our work in some detail. The plan of presentation is the following: Section~\ref{theModel} comprises the model. In Sec.~\ref{RDN} background on random resistor diode networks is provided. Section~\ref{kirchhoffsLaws} discusses Ohm's and Kirchhoff's laws for these networks. Section~\ref{replicaFormalism} introduces a generating function for $M_R$ and explains the involved replica approach.  In Sec.~\ref{fieldTheoreticHamiltonian} we set up a field theoretic Hamiltonian. Section~\ref{RGA} contains the core of our renormalization group analysis. In Sec.~\ref{diagrammaticElements} the diagrammatic elements of our perturbation calculation are presented. Section~\ref{resistanceOfFeynmanDiagrams} is devoted to the real-world interpretation of the Feynman diagrams constituted by these elements. In Sec.~\ref{renormalizationAndScaling} we set up a renormalization group equation. Its solution provides us with the scaling behavior of $M_R$. In Sec.~\ref{comparisonToData} we compare our result for $\phi$ to related results by other authors. Section~\ref{concusions}, finally, contains concluding remarks. Technical details are relegated to Appendices~\ref{app:calculation} and \ref{app:results}.

\section{The model}
\label{theModel}
\subsection{The random resistor diode network}
\label{RDN}
Our approach is based on a model which captures both, IP and DP, namely the random resistor diode network (RDN) introduced by Redner~\cite{red_81&82a,red_83,perc}. A RDN consists of a $d$-dimensional hyper-cubic lattice in which nearest-neighbor sites are connected by a resistor, a positive diode (conducting only in a preferred direction), a negative diode (conducting only opposite to the preferred direction), or an insulator with respective probabilities $p$, $p_{+}$, $p_{-}$, and $q=1-p-p_{+}-p_{-}$. In the three dimensional phase diagram (pictured as a tetrahedron spanned by the four probabilities) one finds a non-percolating and three percolating phases. The percolating phases are isotropic, positively directed, or negatively directed. Between the phases there are surfaces of continuous transitions. All four phases meet along a multicritical line, where $0\leq r:=p_{+}=p_{-}\leq 1/2$ and $p=p_{c}(r)$. On the entire multicritical line, i.e., independently of $r$, one finds the scaling properties of usual isotropic percolation ($r=0$). For the crossover from IP to DP see, e.g., Ref.~\cite{janssen_stenull_2000}.

\subsection{Kirchhoff's laws}
\label{kirchhoffsLaws}
To be specific we choose ${\rm{\bf n}} = 1/\sqrt{d} \left( 1, \dots , 1 \right)$ as the preferred direction. We assume that the bonds $\underline{b}_{\langle i,j \rangle}$ between two nearest neighboring lattice sites $i$ and $j$ are directed so that $\underline{b}_{\langle i,j \rangle} \cdot {\rm{\bf n}} > 0$. We suppose that the directed bonds obey the non-linear Ohm's law
\begin{eqnarray}
\label{avoid}
\sigma_{\underline{b}_{\langle i,j \rangle}} \left( V_{\underline{b}_{\langle i,j \rangle}} \right) V_{\underline{b}_{\langle i,j \rangle}} = I_{\underline{b}_{\langle i,j \rangle}}\, ,
\end{eqnarray}
where $V_{\underline{b}_{\langle i,j \rangle}} = V_j - V_i$ is the voltage drop over the bond between sites $j$ and $i$ and $I_{\underline{b}_{\langle i,j \rangle}}$ denotes the current flowing from $j$ to $i$. In the following we drop the subscript $\langle i,j \rangle$ whenever it is save. The bond conductances $\sigma_{\underline{b}}$ are random variables taking on the values $\sigma$, $\sigma \theta \left( V \right)$, $\sigma \theta \left( -V \right)$, and $0$ with respective probabilities $p$, $p_+$, $p_-$, and $q$. $\sigma$ is a positive constant and $\theta$ denotes the Heaviside function. Note that the diodes are idealized: under forward-bias voltage they behave as ``ohmic'' resistors whereas they are insulating under backward-bias voltage.  To avoid confusion, we point out that the round brackets in Eq.~(\ref{avoid}) contain the functional argument of the bond conductance. Because the bond conductance depends on the voltage drop over that bond only via a Heaviside function and $\mbox{sign} \left( V_{\underline{b}} \right) = \mbox{sign} \left( I_{\underline{b}} \right)$ we may write $\sigma_{\underline{b}} \left( V_{\underline{b}} \right) = \sigma_{\underline{b}} \left( I_{\underline{b}} \right)$.

Assume that an external current $I$ is injected at $x$ and withdrawn at $x^\prime$. It is understood that $x$ and $x^\prime$ are connected. The power dissipated on the network is by definition 
\begin{eqnarray}
P=I \left[ V_x - V_{x^\prime} \right] \, .
\end{eqnarray}
Using Ohm's law it may be expressed entirely in terms of the voltages as
\begin{eqnarray}
\label{power1}
P = R_+(x ,x^\prime)^{-1} \left[ V_x - V_{x^\prime} \right]^2 = \sum_{\underline{b}} \sigma_{\underline{b}} \left( V_{\underline{b}} \right) V_{\underline{b}}^2 = P \left( \left\{ V \right\} \right) \, .
\end{eqnarray}
The sum is taken over all current carrying bonds (the backbone) between $x$ and $x^\prime$ and $\left\{ V \right\}$ denotes the corresponding set of voltages. $R_+(x ,x^\prime)$ stands for the macroscopic resistance when $I$ is inserted at $x$ and withdrawn at $x^\prime$. Similarly one defines $R_-(x ,x^\prime)$ as the macroscopic resistance when $I$ is inserted at $x^\prime$ and withdrawn at $x$. The two quantities are related by $R_+(x ,x^\prime) = R_-(x^\prime ,x)$. From the power one obtains Kirchhoff's first law
\begin{eqnarray}
\label{kirchhoff}
\sum_{\langle j \rangle} \sigma_{\underline{b}_{\langle i,j \rangle}} \left( V_{\underline{b}_{\langle i,j \rangle}} \right) V_{\underline{b}_{\langle i,j \rangle}} = \sum_{\langle j \rangle} I_{\underline{b}_{\langle i,j \rangle}} = - I_i 
\end{eqnarray}
as a consequence of the variation principle
\begin{eqnarray}
\label{variationPrinciple1}
\frac{\partial}{\partial V_i} \left[ \frac{1}{2} P \left( \left\{ V \right\} \right) + \sum_k I_k V_k \right] = 0 \, .
\end{eqnarray}
The summation in Eq.~(\ref{kirchhoff}) extends over the nearest neighbors of $i$ and $I_i$ is given by $I_i = I \left( \delta_{i,x} - \delta_{i,x^\prime} \right)$.

Alternatively to Eq.~(\ref{power1}), the power can be expressed in terms of the currents as
\begin{eqnarray}
\label{power2}
P = R_+ \left( x ,x^\prime \right) I^2 = \sum_{\underline{b}} \sigma_{\underline{b}} \left( I_{\underline{b}} \right)^{-1} I_{\underline{b}}^2 = P \left( \left\{ I \right\} \right) \, ,
\end{eqnarray}
with $\left\{ I \right\}$ denoting the set of currents flowing through the individual bonds. It is understood that $\sigma_{\underline{b}} \left( I_{\underline{b}} \right)^{-1} I_{\underline{b}}^2 =0$ whenever $\sigma_{\underline{b}} \left( I_{\underline{b}} \right) = 0$. Kirchhoff's second law, saying that the voltage drops along closed loops vanish, can be stated in terms of the variation principle
\begin{eqnarray}
\label{variationPrinciple2}
\frac{\partial}{\partial I^{(l)}} P \left( \left\{ I^{(l)} \right\} , I \right) = 0 \, ,
\end{eqnarray}  
i.e., there are no independent loop currents $I^{(l)}$ circulating around a complete set of independent closed loops.

\subsection{Generating function}
\label{replicaFormalism}
Our aim is to calculate the average resistance between two ports $x$ and $x^\prime$. In this section we set up a generating function for it. We demonstrate that this generating function indeed serves its purpose and explain how the average resistance can be extracted from it.

The average resistance we are interested in is precisely defined by
 \begin{eqnarray}
M_R(x ,x^\prime) = \langle \chi_+ (x ,x^\prime) R_+ (x ,x^\prime ) \rangle_C / \langle \chi_+ (x ,x^\prime) \rangle_C \, . 
\end{eqnarray}
$\langle ...\rangle_C$ denotes the average over all configurations $C$ of the 
diluted lattice. $\chi_+ (x ,x^\prime)$ is an indicator function that takes the value one if $x$ and $x^\prime$ are positively connected, i.e., if $I$ can percolate from $x$ to $x^\prime$, and zero otherwise. Note that $\langle \chi_+ (x ,x^\prime) \rangle_C = \langle \chi_- (x^\prime ,x) \rangle_C$ is nothing more than the usual DP correlation function.

We follow an idea by Stephen~\cite{stephen_78} and its generalization to networks of nonlinear resistors by Harris~\cite{harris_87} and exploit correlation functions of
\begin{eqnarray}
\label{defPsi}
\psi_{\vec{\lambda}}(x) = \exp \left( i \vec{\lambda} \cdot \vec{V}_x \right) \ , \quad \vec{\lambda} \neq \vec{0}\, ,
\end{eqnarray}
as generating functions of $M_R$. In writing Eq.~(\ref{defPsi}) we switched to $D$-fold replicated voltages $V_i \to \vec{V_i} = \left( V_i^{(1)}, \cdots , V_i^{(D)} \right)$ and imaginary currents $\lambda_i = i I_i \to \vec{\lambda_i} = \left( \lambda_i^{(1)}, \cdots , \lambda_i^{(D)} \right)$. The correlation functions 
\begin{eqnarray}
G \left( x, x^\prime ,\vec{\lambda} \right) = \left\langle 
\psi_{\vec{\lambda}}(x)\psi_{-\vec{\lambda}}(x^\prime) 
\right\rangle_{\mbox{\scriptsize{rep}}}
\end{eqnarray}
are given by
\begin{eqnarray}
\label{erzeugendeFunktion}
G \left( x, x^\prime ,\vec{\lambda} \right) &=& \bigg\langle Z^{-D} \int \prod_j \prod_{\alpha =1}^D dV_j^{(\alpha )} \exp \bigg[ -\frac{1}{2} P \left( \left\{ \vec{V} \right\} \right)  
\nonumber \\
&+& i \vec{\lambda} \cdot \left( 
\vec{V}_x  - \vec{V}_{x^\prime} \right) \bigg] \bigg\rangle_C \, ,
\end{eqnarray}
where
\begin{eqnarray}
\label{repPower}
P \left( \left\{ \vec{V} \right\} \right) &=& \sum_{\alpha =1}^D P \left( \left\{ V^{(\alpha )} \right\} \right) 
\nonumber \\
&=& \sum_{\alpha =1}^D \bigg\{ \sum_{\underline{b}} \sigma_{\underline{b}} \left(  V_{\underline{b}}^{(\alpha )}  \right) V_{\underline{b}}^{(\alpha )2} + \frac{i\omega}{2} \sum_i \vec{V}^2_i \bigg\}
\end{eqnarray}
and $Z$ is the normalization
\begin{eqnarray}
\label{norm}
Z = \int \prod_{j} dV_{j} \exp \left[ -\frac{1}{2} P \left( \left\{ V \right\} \right) \right] \, .
\end{eqnarray}
Note that we have introduced an additional power term $\frac{i\omega}{2} \sum_i V^2_i$. This is necessary to give the integrals in Eqs.~(\ref{erzeugendeFunktion}) and (\ref{norm}) a well defined meaning. Without this term the integrands depend only on voltage differences and the integrals are divergent. Physically the new term corresponds to grounding each lattice site by a capacitor of unit capacity. The original situation may be restored by taking the limit of vanishing frequency, $\omega \to 0$.

From Eq.~(\ref{erzeugendeFunktion}) the benefit of the replication procedure becomes evident. It provides us with an extra parameter $D$ which we may tune to zero. In this replica limit the normalization denominator $Z^{-D}$ goes to one and hence does not depend on the configurations $C$ anymore. Then the only remaining dependence on $C$ rests in the power $P$ appearing in the exponential in Eq.~(\ref{erzeugendeFunktion}). In the replica limit, therefore, we just have to average this exponential instead of the entire right hand side of Eq.~(\ref{erzeugendeFunktion}). This average then provides us with an effective power or Hamiltonian which serves as vantage point for all further calculations. The effective Hamiltonian will be discussed in Sec.~\ref{fieldTheoreticHamiltonian}.

Now we come back to the role of Eq.~(\ref{erzeugendeFunktion}) as a generating function. The integrations in Eqs.~(\ref{erzeugendeFunktion}) and (\ref{norm}) are not Gaussian due to the $\theta$ functions. Therefore we employ the saddle point method. The saddle point equation is nothing more than the variation principle stated in Eq.~(\ref{variationPrinciple1}). Thus, the maximum of the integrand is determined by the solution of the circuit equations (\ref{kirchhoff}). Provided that the condition $I^2 \gg \sigma$ holds, we obtain
\begin{eqnarray}
\label{GenFkt}
G \left( x, x^\prime ,\vec{\lambda} \right) = \left\langle \exp \left[ - 
\frac{\vec{\lambda}^2}{2} R_+ \left( x,x^\prime \right)  \right] \right\rangle_C 
\end{eqnarray}
up to an unimportant multiplicative constant which goes to one in the limit $D \to 0$. Upon expanding $G$ around $\vec{\lambda}^2 =0$, 
\begin{eqnarray}
\label{expansionOfG}
G \left( x, x^\prime ,\vec{\lambda} \right) = \left\langle \chi_+ ( x,  x^\prime ) \right\rangle_C \left( 1 - \frac{\vec{\lambda}^2 }{2} M_R ( x, x^\prime ) + \cdots 
\right) \, ,
\end{eqnarray}
we learn that the average resistance may be calculated via
\begin{eqnarray}
\label{shit}
M_R ( x, x^\prime ) = \left\langle \chi_+ ( x,  x^\prime ) \right\rangle_C^{-1} \frac{\partial}{\partial \left( - \vec{\lambda}^2/2 \right)} G \left( x, x^\prime ,\vec{\lambda} \right) \Big|_{\vec{\lambda}^2 =0} \, ,
\end{eqnarray}
i.e., $G$ is indeed a generating function for $M_R$.

At this point a comment on the nature of $\vec{\lambda}$ is appropriate. We work near the limit when all the components of $\vec{\lambda}$ are equal and continue to large imaginary values. Accordingly we set~\cite{harris_87}
\begin{eqnarray}
\label{lambdaChoice}
\lambda^{(\alpha )} = i \lambda_0 + \xi^{(\alpha )}
\end{eqnarray}
with real $\lambda_0$ and $\xi^{(\alpha )}$, $\sum_{\alpha =1}^D \xi^{(\alpha )} = 0$. The saddle point approximation in Eq.~(\ref{GenFkt}) may be justified by demanding
\begin{eqnarray}
\label{cond1}
\left| \lambda_0 \right| \gg 1 \, . 
\end{eqnarray}
On the other hand one has
\begin{eqnarray}
 \vec{\lambda}^2 = - D  \lambda_0^2 + \vec{\xi}^2 \, .
\end{eqnarray}
Thus, one can justify the expansion in Eq.~(\ref{expansionOfG}) by invoking the conditions
\begin{eqnarray}
\label{cond2}
\lambda_0^{2} \ll D^{-1} \quad \mbox{and} \quad \vec{\xi}^2 \ll 1 \, .
\end{eqnarray}
Note that the replica limit $D\to 0$ allows for a simultaneous fulfillment of the conditions (\ref{cond1}) and (\ref{cond2}).

\subsection{Field theoretic Hamiltonian}
\label{fieldTheoreticHamiltonian}
In this section we device a field theoretic Hamiltonian for resistor diode percolation. Starting point will be the effective Hamiltonian announced in Sec.~\ref{replicaFormalism}. First of all, however, we need to take care of some further regularization issues.

Since infinite voltage drops between different clusters may occur, it is not guaranteed that $Z$ stays finite, i.e., the limit $\lim_{D \to 0}{Z^D}$ is not well defined. This problem can be regularized by switching to voltage variables $\vec{\vartheta}$ taking discrete values  on a $D$-dimensional torus which we refer to as the replica space. The voltages are discretized by setting $\vec{\vartheta} = \Delta \vartheta \vec{k}$, where $\Delta \vartheta = \vartheta_M /M$ is the gap between successive voltages, $\vartheta_M$ is a voltage cutoff, $\vec{k}$ is a $D$-dimensional integer, and $M$ a positive integer. The components of $\vec{k}$ are restricted to $-M < k^{(\alpha)} \leq M$ and periodic boundary conditions are realized by equating $k^{(\alpha )}=k^{(\alpha )} \mbox{mod} (2M)$. The continuum may be restored by taking $\vartheta_M \to \infty$ and $\Delta \vartheta \to 0$. By setting $\vartheta_M = \vartheta_0 M$, $M=m^2$, and, respectively, $\Delta \vartheta = \vartheta_0 /m$, the two limits can be taken simultaneously via $m \to \infty$. Note that the limit $D \to 0$ has to be taken before any other limit. Since the voltages and $\vec{\lambda}$ are conjugated variables, $\vec{\lambda}$ is affected by the discretization as well:
\begin{eqnarray}
\vec{\lambda} = \Delta \lambda \, \vec{l} \ , \ \Delta \lambda \, \Delta \vartheta = \pi /M \, ,
\end{eqnarray}
where $\vec{l}$ is a $D$-dimensional integer taking the same values as $\vec{k}$. This choice guarantees that the completeness and orthogonality relations
\begin{subequations}
\label{complete}
\begin{eqnarray}
\frac{1}{(2M)^D} \sum_{\vec{\vartheta}} \exp \left( i \vec{\lambda} \cdot \vec{\vartheta} \right) = \delta_{\vec{\lambda} ,\vec{0} 
\hspace{0.15em}\mbox{\scriptsize{mod}}(2M \Delta \lambda) }
\end{eqnarray}
and
\begin{eqnarray}
\frac{1}{(2M)^D} \sum_{\vec{\lambda}} \exp \left( i \vec{\lambda} \cdot \vec{\vartheta} \right) = \delta_{\vec{\vartheta} ,\vec{0} 
\hspace{0.15em}\mbox{\scriptsize{mod}}(2M \Delta \vartheta)}
\end{eqnarray}
\end{subequations}
do hold. Equation~(\ref{complete}) provides us with a Fourier transform between the $\vec{\vartheta}$- and $\vec{\lambda}$-tori.

Now we revisit Eq.~(\ref{erzeugendeFunktion}). The replica limit provides us with the effective Hamiltonian
\begin{eqnarray}
\label{effHamil1}
H_{\mbox{\scriptsize{rep}}} &=&  - \ln \left\langle  \exp \left[ - \frac{1}{2} P \left( \left\{ \vec{\vartheta} \right\} \right) \right] \right\rangle_C 
\nonumber \\
&=& - \ln \bigg\{ \int_0^\infty \prod_{\underline{b}} d \sigma_{\underline{b}} \, f \left( \sigma_{\underline{b}} \right) \exp \left[ - \frac{1}{2} P \left( \left\{ \vec{\vartheta} \right\} \right) \right] \bigg\} \, ,
\end{eqnarray}
where the distribution function of the bond conductances is given by
\begin{eqnarray}
 f \left( \sigma_{\underline{b}} \right) &=& p \, \delta \left[ \sigma_{\underline{b}} - \sigma \right] +  p_+ \, \delta \left[ \sigma_{\underline{b}} - \sigma \theta (V) \right] 
\nonumber \\
&+& p_- \, \delta \left[ \sigma_{\underline{b}} - \sigma \theta (-V) \right] + q \, \delta \left[ \sigma_{\underline{b}} \right]
\, .
\end{eqnarray}
The $\delta$-functions are interpreted so that $\int_0^\infty dx \, \delta \left[ x \right] =1$. Carrying out the average yields
\begin{eqnarray}
\label{effHamil2}
H_{\mbox{\scriptsize{rep}}} = - \sum_{\underline{b}} K \left( \vec{\vartheta}_{\underline{b}} \right) - \frac{i\omega}{2} \sum_i \vec{\vartheta}_{i}^2 \, ,
\end{eqnarray}
where we have introduced
\begin{eqnarray}
\label{kern1}
K \left( \vec{\vartheta}\right) &=& \ln \bigg\{ q + p \exp \left[ - \frac{\sigma}{2} \vec{\vartheta}^2 \right]
+ p_+ \prod_{\alpha =1}^D \exp \left[ - \frac{\sigma}{2} \theta \left( \vartheta^{(\alpha )} \right) \vartheta^{(\alpha )2} \right] 
\nonumber \\
&+& p_- \prod_{\alpha =1}^D \exp \left[ - \frac{\sigma}{2} \theta \left( - \vartheta^{(\alpha )} \right) \vartheta^{(\alpha )2} \right] \bigg\} \, .
\end{eqnarray}
Now recall our choice for $\vec{\lambda}$ in Eq.~(\ref{lambdaChoice}). Since $\vec{\lambda}$ and $\vec{\vartheta}$ are related via Ohm's law Eq.~(\ref{avoid}), we have to choose $\vec{\vartheta}$ consistently:
\begin{eqnarray}
\vartheta^{(\alpha )} = \vartheta_0 + \zeta^{(\alpha )}
\end{eqnarray}
with real $\vartheta_0$ and $\zeta^{(\alpha )}$, $\sum_{\alpha =1}^D \zeta^{(\alpha )} = 0$. We impose the conditions
\begin{eqnarray}
\left| \vartheta_0 \right| \gg 1 \ , \quad \vartheta_0^2 \ll D^{-1} \ , \quad \vec{\zeta}^2 \ll 1 \, .
\end{eqnarray}
Under these conditions we can write
\begin{eqnarray}
\label{kern2}
K \left( \vec{\vartheta}\right) &=& \ln \bigg\{ q + p \exp \left[ - \frac{\sigma}{2} \vec{\vartheta}^2 \right]
+ p_+ \left[ \theta \left( -  \vartheta_0 \right) + \theta \left( \vartheta_0 \right) \exp  \left( - \frac{\sigma}{2} \vec{\vartheta}^2 \right) \right] 
\nonumber \\
&+& p_- \left[ \theta \left( \vartheta_0 \right) + \theta \left( - \vartheta_0 \right) \exp \left( - \frac{\sigma}{2} \vec{\vartheta}^2 \right) \right] \bigg\} \, .
\end{eqnarray}
By doing a little algebra we recast Eq.~(\ref{kern2}) as
\begin{eqnarray}
\label{kern3}
K \left( \vec{\vartheta}\right) = \theta \left( \vartheta_0 \right) K_+ \left( \vec{\vartheta}\right) + \theta \left( - \vartheta_0 \right) K_- \left( \vec{\vartheta}\right) \, .
\end{eqnarray}
Here we dropped a term
\begin{eqnarray}
\theta \left( \vartheta_0 \right) \ln \left[ 1 - p - p_+ \right] + \theta \left( - \vartheta_0 \right) \ln \left[ 1 - p - p_- \right] \, ,
\end{eqnarray}
which is not depending on the bond conductances. Moreover, we defined
\begin{eqnarray}
\label{defKpm}
K_\pm \left( \vec{\vartheta}\right) = \ln \left[ 1 - \frac{p + p_\pm}{ 1 - p - p_\pm} \exp \left( - \frac{\sigma}{2} \vec{\vartheta}^2 \right) \right] \, .
\end{eqnarray}
Note that $K_\pm \left( \vec{\vartheta} \right)$ are exponentially decreasing functions in replica space with a decay rate proportional to $\sigma^{-1}$. Next we expand $K \left( \vec{\vartheta}\right)$ in terms of the $\psi_{\vec{\lambda}}(i) = \exp \left( i \vec{\lambda} \cdot \vec{\vartheta}_i \right)$,
\begin{eqnarray}
\label{kern4}
K \left( \vec{\vartheta}_{\underline{b}} \right) &=& \frac{1}{\left( 2M \right)^D} \sum_{\vec{\lambda}} \sum_{\vec{\vartheta}} \exp \left[ i \vec{\lambda} \cdot \left( \vec{\vartheta}_{\underline{b}} - \vec{\vartheta} \right) \right] K \left( \vec{\vartheta} \right)
\nonumber \\
&=& \sum_{\vec{\lambda} \neq \vec{0}} \psi_{\vec{\lambda}} \left( i \right) \psi_{-\vec{\lambda}} \left( j \right) 
\nonumber \\
&\times& \left[  \theta \left( \lambda_0 \right)\widetilde{K}_+ \left( \vec{\lambda} \right) + \theta \left( -\lambda_0 \right)\widetilde{K}_- \left( \vec{\lambda} \right) \right] \, ,
\end{eqnarray}
where we have exploited that $\theta \left( \vartheta_0 \right) = \theta \left( \lambda_0 \right)$. $\widetilde{K}_\pm \left( \vec{\lambda} \right)$ stands for the Fourier transform of $K_\pm \left( \vec{\vartheta}\right)$, 
\begin{eqnarray}
\widetilde{K}_\pm \left( \vec{\lambda} \right) = \frac{1}{\left( 2M \right)^D} \sum_{\vec{\vartheta}} \exp \left[ i \vec{\lambda} \cdot \vec{\vartheta} \right] K \left( \vec{\vartheta} \right) \, .
\end{eqnarray}
Because $K_\pm \left( \vec{\vartheta}\right)$ are rotationally invariant functions in replica space, their Fourier transforms can be Taylor expanded as
\begin{eqnarray}
\widetilde{K}_\pm \left( \vec{\lambda} \right) = \tau_\pm + w_\pm \vec{\lambda}^2 + \cdots \, ,
\end{eqnarray}
where $\tau_\pm$ and $w_\pm \sim \sigma^{-1}$ are expansion coefficients depending on $p$ and $p_\pm$. These expansion coefficients satisfy $\tau_\pm (p, p_+ ,p_-) = \tau_\mp (p, p_- ,p_+)$ and $w_\pm (p, p_+ ,p_-) = w_\mp (p, p_- ,p_+)$. Since $\widetilde{K}_\pm \left( \vec{\lambda} \right)$ depends only on the square modulus of $\vec{\lambda}$ it is easy to decompose $K \left( \vec{\vartheta}_{\underline{b}} \right)$ into two parts, one being even and the other being odd under $\vec{\lambda} \to -\vec{\lambda}$:
\begin{eqnarray}
\label{kern5}
K \left( \vec{\vartheta}_{\underline{b}} \right) 
&=& \sum_{\vec{\lambda} \neq \vec{0}} \psi_{\vec{\lambda}} \left( i \right) \psi_{-\vec{\lambda}} \left( j \right) 
\bigg\{ \frac{1}{2} \left[ \widetilde{K}_+ \left( \vec{\lambda} \right) + \widetilde{K}_- \left( \vec{\lambda} \right) \right] 
\nonumber \\
&& + \frac{1}{2} \left[ \theta \left( \lambda_0 \right) - \theta \left( -\lambda_0 \right) \right] \left[ \widetilde{K}_+ \left( \vec{\lambda} \right) - \widetilde{K}_- \left( \vec{\lambda} \right) \right] \bigg\} \, .
\end{eqnarray}
Now we insert Eq.~(\ref{kern5}) into Eq.~(\ref{effHamil2}). We also carry out a gradient expansion in position space which is justified because the interaction is short ranged not only in replica but also in position space. We find 
\begin{eqnarray}
\label{effHamil3}
H_{\mbox{\scriptsize{rep}}} &=&  - \sum_{\vec{\lambda} \neq \vec{0}} \sum_{i , \underline{b}_i} \bigg\{ \frac{1}{2} \left[ \widetilde{K}_+ \left( \vec{\lambda} \right) + \widetilde{K}_- \left( \vec{\lambda} \right) \right] 
\nonumber \\
&\times&
\psi_{-\vec{\lambda}} \left( i \right) \left[ 1  + \frac{1}{2} \left( \underline{b}_i \cdot \nabla \right)^2 + \cdots \right] \psi_{\vec{\lambda}} \left( i \right)
\nonumber \\
&+&  \frac{1}{2} \left[ \theta \left( \lambda_0 \right) - \theta \left( -\lambda_0 \right) \right] \left[ \widetilde{K}_+ \left( \vec{\lambda} \right) - \widetilde{K}_- \left( \vec{\lambda} \right) \right] 
\nonumber \\
&\times&
\psi_{-\vec{\lambda}} \left( i \right) \left[ \underline{b}_i \cdot \nabla + \cdots \right] \psi_{\vec{\lambda}} \left( i \right) \, .
\end{eqnarray}

We proceed with the usual coarse graining step and replace the 
$\psi_{\vec{\lambda}} \left( i \right)$ by order parameter fields $\psi_{\vec{\lambda}} \left( {\rm{\bf x}} \right)$ which inherit the constraint $\vec{\lambda} \neq \vec{0}$. We model the corresponding field theoretic Hamiltonian $\mathcal{H}$ in the spirit of Landau as a mesoscopic free energy and introduce the Landau-Ginzburg-Wilson type functional
\begin{eqnarray}
\label{hamiltonian}
{\mathcal{H}} &=& \int d^dx \bigg\{ \frac{1}{2} \sum_{\vec{\lambda} \neq \vec{0}} \psi_{-\vec{\lambda}} \left( {\rm{\bf x}} \right) \Big[  \tau - \nabla^2 + w \vec{\lambda}^2 
\nonumber \\
&+& \left( \theta \left( \lambda_0 \right) - \theta \left( -\lambda_0 \right) \right) {\rm{\bf v}} \cdot \nabla \Big] \psi_{\vec{\lambda}} \left( {\rm{\bf x}} \right)
\nonumber \\
&+& \frac{g}{6} \sum_{\vec{\lambda}, \vec{\lambda}^\prime  , \vec{\lambda} + \vec{\lambda}^\prime \neq \vec{0}} \psi_{-\vec{\lambda}} \left( {\rm{\bf x}} \right) \psi_{-\vec{\lambda}^\prime} \left( {\rm{\bf x}} \right) \psi_{\vec{\lambda} + \vec{\lambda}^\prime} \left( {\rm{\bf x}} \right) 
\nonumber \\
&+& i \omega \sum_{\vec{\lambda} \neq \vec{0}} \nabla^2_{\vec{\lambda}} \psi_{\vec{\lambda}} \left( {\rm{\bf x}} \right) \bigg\} \, ,
\end{eqnarray}
where we have discarded all terms irrelevant in the renormalization group sense. The parameter $\tau$ is the coarse grained relative of $\tau_+ + \tau_-$. It specifies the ``distance'' from the critical surface under consideration. $w \sim \sigma^{-1}$ is the coarse grained analog of $w_+ + w_-$. The vector ${\rm{\bf v}}$ lies in the preferred direction, ${\rm{\bf v}} = v {\rm{\bf n}}$. $\tau$, $w$, and $v$ depend on the three probabilities $p$, $p_+$, and $p_-$. $v$ is zero if $p_+ = p_-$. We will see as we go along that our Hamiltonian ${\mathcal{H}}$ describes in the limit $w\to 0$ the usual purely geometric DP. Indeed ${\mathcal{H}}$ leads for $w\to 0$ to exactly the same perturbation series as obtained in~\cite{cardy_sugar_80,janssen_81,janssen_2000}.

\section{Renormalization group analysis}
\label{RGA}
In this section we utilize ${\mathcal{H}}$ and calculate the generating function $G ( {\rm{\bf x}}, {\rm{\bf x}}^\prime , \vec{\lambda} )$ by employing field theory augmented by renormalization. For background on these methods we refer to Ref.~\cite{amit_zinn-justin}. We perform a diagrammatic perturbation calculation up to two-loop order. In the remainder we drop the regularization term proportional to $\omega$ for simplicity. 

\subsection{Diagrammatic elements}
\label{diagrammaticElements}
Instead of working with ${\mathcal{H}}$ directly, we recast ${\mathcal{H}}$ into the form of a dynamic functional~\cite{janssen_dynamic,deDominicis&co,janssen_92}. This strategy is convenient because it simplifies the following calculations from the onset. Assuming $v \neq 0$ we introduce new variables by setting
\begin{eqnarray}
\label{subst}
x_\parallel = {\rm{\bf n}} \cdot {\rm{\bf x}} = v \rho t \, , \quad 
\psi = |v|^{-1/2} \, s \, , \quad
g = |v|^{1/2} \, \overline{g} \, .
\end{eqnarray}
On substituting Eq.~(\ref{subst}) into Eq.~(\ref{hamiltonian}) we obtain the dynamic functional
\begin{eqnarray}
\label{dynFktnal}
{\mathcal{J}} &=& \int d^{d_\perp}x_\perp  \, dt \bigg\{ \frac{1}{2} \sum_{\vec{\lambda} \neq \vec{0}} s_{-\vec{\lambda}} \left( {\rm{\bf x}}_\perp , t \right) \Big[ \rho \left( \tau - \nabla^2_\perp + w \vec{\lambda}^2 \right) 
\nonumber \\
&+& \left( \theta \left( \lambda_0 \right) - \theta \left( -\lambda_0 \right) \right) \frac{\partial}{\partial t} \Big] s_{\vec{\lambda}} \left( {\rm{\bf x}}_\perp , t  \right)
\nonumber \\
&+& \frac{\rho \overline{g}}{6} \sum_{\vec{\lambda}, \vec{\lambda}^\prime  , \vec{\lambda} + \vec{\lambda}^\prime \neq \vec{0}} s_{-\vec{\lambda}} \left( {\rm{\bf x}}_\perp , t  \right) s_{-\vec{\lambda}^\prime} \left( {\rm{\bf x}}_\perp , t  \right) s_{\vec{\lambda} + \vec{\lambda}^\prime} \left( {\rm{\bf x}}_\perp , t  \right) \bigg\} \, ,
\end{eqnarray}
where $d_\perp = d-1$. In Eq.~(\ref{dynFktnal}) we dropped a term containing a second derivative with respect to $t$ because it is irrelevant compared to the retained term containing $\partial / \partial t$. 

From Eq.~(\ref{dynFktnal}) we gather the diagrammatic elements contributing to our renormalization group improved perturbation calculation. Dimensional analysis shows that the coupling constant $\overline{g}$ has the naive dimension $(4 - d_\perp )/2$, i.e., $d = d_\perp +1 = 5$ is the upper critical dimension. The Gaussian propagator $G \left( {\rm{\bf x}}_\perp , t , \vec{\lambda} \right)$ is determined by the equation of motion
\begin{eqnarray}
&& \left\{ \rho \left( \tau - \nabla^2 + w \vec{\lambda}^2 \right) + \left( \theta \left( \lambda_0 \right) - \theta \left( -\lambda_0 \right) \right) \frac{\partial}{\partial t} \right\} 
\nonumber \\
&&\, \times G \left( {\rm{\bf x}}_\perp , t , \vec{\lambda} \right) = \delta \left( {\rm{\bf x}}_\perp \right) \delta \left( t \right) .
\end{eqnarray}
Solving the equation of motion is straightforward. For the Fourier transform $\widetilde{G} \left( {\rm{\bf p}} , t , \vec{\lambda} \right)$ of $G \left( {\rm{\bf x}}_\perp , t , \vec{\lambda} \right)$ we obtain
\begin{eqnarray}
\widetilde{G} \left( {\rm{\bf p}} , t , \vec{\lambda} \right) = \widetilde{G}_+ \left( {\rm{\bf p}} , t , \vec{\lambda} \right) + \widetilde{G}_- \left( {\rm{\bf p}} , t , \vec{\lambda} \right) \, ,
\end{eqnarray}
where ${\rm{\bf p}}$ is the momentum conjugate to ${\rm{\bf x}}_\perp$ and
\begin{eqnarray}
\label{defGpm}
\widetilde{G}_\pm \left( {\rm{\bf p}} , t , \vec{\lambda} \right) = \theta \left( \pm t \right) \theta \left( \pm \lambda_0 \right) \exp \left[ \mp t \rho \left( \tau + {\rm{\bf p}}^2 + w \vec{\lambda}^2 \right) \right] \left( 1 - \delta_{\vec{\lambda}, \vec{0}} \right) \, .
\end{eqnarray}
For the perturbation expansion it is sufficient to keep either $\widetilde{G}_+ \left( {\rm{\bf p}} , t , \vec{\lambda} \right)$ or $\widetilde{G}_- \left( {\rm{\bf p}} , t , \vec{\lambda} \right)$ and hence we discard $\widetilde{G}_- \left( {\rm{\bf p}} , t , \vec{\lambda} \right)$. 

\subsection{Resistance of Feynman diagrams}
\label{resistanceOfFeynmanDiagrams}
\begin{figure}
\epsfxsize=10cm
\begin{center}
\epsffile{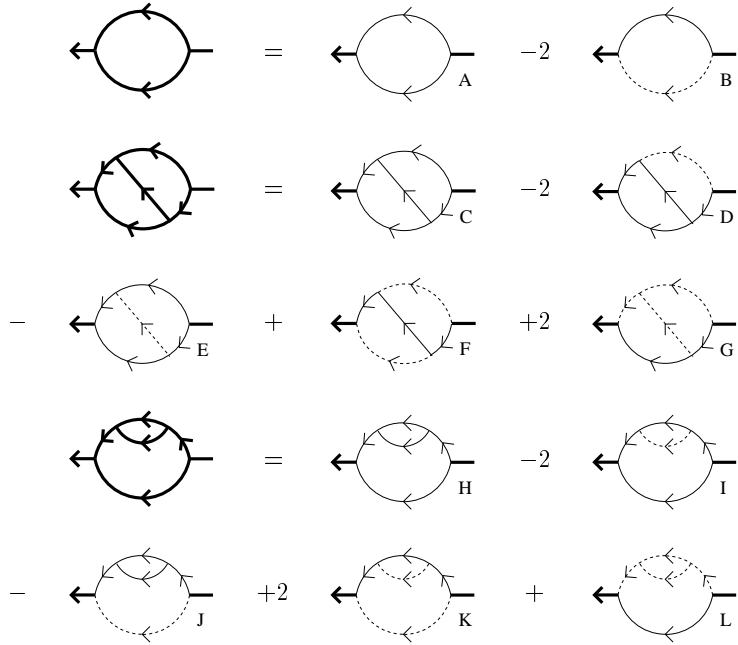}
\end{center}
\caption[]{\label{fig1}Here we list the decomposition of the primary two leg diagrams (bold) into conducting diagrams composed of conducting (light) and insulating (dashed) propagators. The listing extends up to two-loop order. Note that the conducting diagrams inherit their combinatorial factor from their bold diagram. For example, the diagrams A and B introduced below have to be calculated with the same combinatorial factor, namely $\frac{1}{2}$.}
\end{figure}

We may rewrite $\widetilde{G}_+ \left( {\rm{\bf p}} , t , \vec{\lambda} \right)$ as
\begin{eqnarray}
\label{decoGp}
\widetilde{G}_+ \left( {\rm{\bf p}} , t , \vec{\lambda} \right) &=& \theta \left( t \right) \theta \left( \lambda_0 \right) \exp \left[ - t \rho \left( \tau + {\rm{\bf p}}^2 + w \vec{\lambda}^2 \right) \right] 
\nonumber \\
&-& \theta \left( t \right) \exp \left[ - t \rho \left( \tau + {\rm{\bf p}}^2 \right) \right] \delta_{\vec{\lambda}, \vec{0}} \, ,
\end{eqnarray}
i.e., the principal propagator $\widetilde{G}_+ \left( {\rm{\bf p}} , t , \vec{\lambda} \right)$ decomposes into a propagator carrying $\vec{\lambda}$'s (conducting) and one not carrying $\vec{\lambda}$'s (insulating). This allows for a schematic decomposition of the principal diagrams into sums of conducting diagrams consisting of conducting and insulating propagators. In Fig.~\ref{fig1} we list the result of the decomposition procedure.

As in our previous work on transport in IP~\cite{stenull_janssen_oerding_99,janssen_stenull_oerding_99,janssen_stenull_99,stenull_2000,stenull_janssen_2000a}, the conducting diagrams may be viewed as resistor networks themselves with conducting propagators corresponding to conductors and insulating propagators corresponding to open bonds. The ``times'' $t$ appearing in the conducting propagators correspond to resistances and the replica variables $i\vec{\lambda}$ to currents. The replica currents are conserved in each vertex and we may write for each edge $i$ of a diagram, $\vec{\lambda}_i = \vec{\lambda}_i \left( \vec{\lambda} , \left\{ \vec{\kappa} \right\} \right)$, where $\vec{\lambda}$ is an external current and $\left\{ \vec{\kappa} \right\}$ denotes a complete set of independent loop currents. The $\vec{\lambda}$-dependent part of a diagram can be expressed in terms of its power $P$:
\begin{eqnarray}
\exp \left[ - \rho \, w \sum_{i} t_i \vec{\lambda}_i^2 \right] = \exp \left[ \rho \, w P \left( \vec{\lambda} , \left\{ \vec{\kappa} \right\} \right) \right] \, .
\end{eqnarray} 

The real-world interpretation suggests an alternative way of computing the Feynman diagrams. To evaluate the sums over independent loop currents,
\begin{eqnarray}
\label{toEvaluate}
\sum_{\left\{ \vec{\kappa} \right\}} \exp \left[ \rho \, w P \left( \vec{\lambda} , \left\{ \vec{\kappa} \right\} \right) \right] \, ,
\end{eqnarray}
we employ the saddle point method under the conditions discussed at the end of Sec.~\ref{replicaFormalism}. Note that the saddle point equation is nothing more than the variation principle stated in Eq.~(\ref{variationPrinciple2}). Thus, solving the saddle point equations is equivalent to determining the total resistance $R \left( \left\{ t_i \right\} \right)$ of a diagram, and the saddle point evaluation of (\ref{toEvaluate}) yields
\begin{eqnarray}
\exp \left[ - R \left(  \left\{ t_i \right\} \right) \rho \, w \vec{\lambda}^2 \right] \, , 
\end{eqnarray}
where we have omitted once more multiplicative factors which go to one for $D \to0$. A completion of squares in the momenta renders the momentum integrations straightforward. Equally well we can use the saddle point method which is exact here since the momentum dependence is purely quadratic. After an expansion for small $w \vec{\lambda}^2$ all diagrammatic contributions are of the form
\begin{eqnarray}
\label{expansionOfDiagrams}
I \left( {\rm{\bf p}}^2 , t, \vec{\lambda}^2 \right) &=& I_P \left( {\rm{\bf p}}^2 , t \right) - I_W 
\left( {\rm{\bf p}}^2 , t \right) \rho \, w  \vec{\lambda}^2 + \cdots
\nonumber \\
&=& \int_0^\infty \prod_i dt_i \left[ 1 - R \left(  \left\{ t_i \right\} 
\right) \rho \, w \vec{\lambda}^2 + \cdots \right] D \left( {\rm{\bf p}}^2,t ; 
\left\{ t_i \right\} \right) \, ,
\end{eqnarray}
where $D \left( {\rm{\bf p}}^2, t ; \left\{ t_i \right\} \right)$ is a typical integrand as known from the field theory of DP~\cite{cardy_sugar_80,janssen_81,janssen_2000}. Concrete examples for calculations of this type are given in Appendix~\ref{app:calculation}.

\subsection{Renormalization and scaling}
\label{renormalizationAndScaling}
We proceed with standard techniques of renormalized field theory~\cite{amit_zinn-justin}. The ultraviolet divergences occurring in the diagrams can be absorbed by dimensional regularization. We employ the renormalization scheme
\begin{subequations}
\label{renorScheme}
\begin{eqnarray}
s \to {\mathaccent"7017 s} = Z^{1/2} s \ ,&\quad&
\tau \to {\mathaccent"7017 \tau} = Z^{-1} Z_{\tau} \tau \, ,
\\
w \to {\mathaccent"7017 w} = Z^{-1} Z_{w} w \ , &\quad&
\rho \to {\mathaccent"7017 \rho} = Z^{-1} Z_{\rho} \rho \, ,
\\
\overline{g} \to \mathaccent"7017{\overline{g}} &=& Z^{-1/2} Z^{-1}_\rho Z_u^{1/2} G_\epsilon^{-1/2} u^{1/2} 
\mu^{\epsilon /2} \, ,
\end{eqnarray}
\end{subequations}
where $\epsilon = 4-d_\perp$ and $\mu$ is the usual inverse length scale. The factor $G_\epsilon = (4\pi )^{-d_\perp/2}\Gamma (1 + \epsilon /2)$, with $\Gamma$ denoting the Gamma function, is introduced for convenience. $Z$, $Z_\tau$, $Z_\rho$, and $Z_u$ are the usual DP Z-factors, which can be found to second order in $\epsilon$ in the literature~\cite{janssen_81,janssen_2000}. Thus, $Z_w$ remains to be determined. In a two-loop calculation we compute the $\vec{\lambda}$-dependent parts of the diagrams displayed in Fig.~1 in dimensional regularization. The corresponding results are listed in App.~\ref{app:results}. From the two-point vertex function, constituted by the inverse Gaussian propagator and the diagrams given in Fig.~1, we extract $Z_w$ by minimal subtraction which gives
\begin{eqnarray}
Z_w = 1 + \frac{3 \, u}{8 \, \epsilon} + \frac{9 \, u^2}{64 \, \epsilon} \left[ \frac{5}{2 \, \epsilon} - \frac{71}{72} + \frac{35}{36} \ln \left( \frac{4}{3} \right) \right] \, .
\end{eqnarray}
There are two points which we would like to emphasize. First, non-primitive divergencies cancel each other which is an indication for the renormalizability of the perturbation expansion. Second, for $w \to 0$ the decomposition in Fig.~1 collapses to the usual DP diagrams with the correct fore-factors, i.e., we indeed retrieve the perturbation series for purely geometric DP. 

The unrenormalized theory has to be independent of the length scale $\mu^{-1}$ introduced by renormalization. In particular, the unrenormalized connected $N$ point correlation functions must be independent of $\mu$, i.e.,  
\begin{eqnarray}
\label{independence}
\mu \frac{\partial}{\partial \mu} {\mathaccent"7017 G}_N \left( \left\{ {\rm{\bf x}}_\perp , {\mathaccent"7017 \rho} t , {\mathaccent"7017 w} \vec{\lambda}^2 \right\} ; {\mathaccent"7017 \tau}, \mathaccent"7017{\overline{g}} \right) = 0
\end{eqnarray}
for all $N$. Eq.~(\ref{independence}) translates via the Wilson functions defined by
\begin{subequations}
\begin{eqnarray}
\label{wilson}
\gamma_{...} \left( u \right) &=& \mu \frac{\partial }{\partial \mu} \ln Z_{...}  \bigg|_0 \, ,
\\
\beta \left( u \right) &=& \mu \frac{\partial u}{\partial \mu} \bigg|_0 = \left( - \epsilon + \gamma + 2 \gamma_\rho - \gamma_u \right) u \, ,
\\ 
\kappa \left( u \right) &=& \mu \frac{\partial
\ln \tau}{\partial \mu}  \bigg|_0 = \gamma_\rho - \gamma_\tau \, ,
 \\
\zeta_w \left( u \right) &=& \mu \frac{\partial \ln w}{\partial \mu}  \bigg|_0 = \gamma_\rho - \gamma_w \, ,
\\
\zeta_\rho \left( u \right) &=& \mu \frac{\partial \ln \rho}{\partial \mu}  \bigg|_0 = \gamma - \gamma_\rho \, ,
\end{eqnarray}
\end{subequations}
where the bare quantities are kept fix while taking the derivatives, into the Gell--Mann-Low renormalization group equation
\begin{eqnarray}
&&\left[ \mu \frac{\partial }{\partial \mu} + \beta \frac{\partial }{\partial u} + 
\tau \kappa \frac{\partial }{\partial \tau} + w \zeta_w \frac{\partial }{\partial w} + \rho \zeta_\rho \frac{\partial }{\partial \rho} +
\frac{N}{2} \gamma \right] 
\nonumber \\
&&\, \times
G_N \left( \left\{ {\rm{\bf x}}_\perp , \rho t  ,w \vec{\lambda}^2 \right\} ; \tau, u, \mu \right) = 0 \, .
\end{eqnarray}

The renormalization group equation is solved by the method of characteristics. At the infrared stable fixed point $u^\ast$, determined by $\beta \left( u^\ast \right) = 0$, the solution reads
\begin{eqnarray}
\label{SolOfRgg}
&&G_N \left( \left\{ {\rm{\bf x}}_\perp , \rho t ,w \vec{\lambda}^2 \right\} ; \tau, u, \mu \right) 
\nonumber \\
&&\, = l^{\gamma^\ast N/2} G_N \left( \left\{ l{\rm{\bf x}}_\perp ,l^{\zeta_\rho^\ast}\rho t , l^{\zeta_w^\ast}w \vec{\lambda}^2 \right\} ; l^{\kappa^\ast}\tau , u^\ast, l \mu \right) \, , 
\end{eqnarray}
where $\gamma^\ast = \gamma \left( u^\ast \right)$, $\kappa^\ast = \kappa \left( u^\ast \right)$, $\zeta_\rho^\ast = \zeta_\rho \left( u^\ast \right)$ , and $\zeta_w^\ast = \zeta_w \left( u^\ast \right)$.

To derive a scaling relation for the correlation functions, a dimensional analysis remains to be done. It yields
\begin{eqnarray}
\label{dimAna}
&&G_N \left( \left\{ {\rm{\bf x}}_\perp ,\rho t ,w \vec{\lambda}^2 \right\} ; \tau, u, \mu \right) 
\nonumber \\
&&\, = 
\mu^{d_\perp N/2} G_N \left( \left\{ \mu {\rm{\bf x}}_\perp , \mu^2 \rho t , \mu^{-2}w \vec{\lambda}^2 \right\} ; \mu^{-2}\tau , u, 1 \right) \, . 
\end{eqnarray}
Equation~(\ref{SolOfRgg}) in conjunction with Eq.~(\ref{dimAna}) now gives
\begin{eqnarray}
\label{scaling}
&&G_N \left( \left\{ {\rm{\bf x}}_\perp , \rho t ,w \vec{\lambda}^2 \right\} ; \tau, u, \mu \right)
\nonumber \\
&&\, = 
l^{(d_\perp +\eta)N/2} G_N \left( \left\{ l{\rm{\bf x}}_\perp , l^z \rho t , l^{-\phi/\nu_\perp}w \vec{\lambda}^2 \right\} ; 
l^{-1/\nu_\perp}\tau , u^\ast, \mu \right) \, .
\end{eqnarray}
Here
\begin{subequations}
\begin{eqnarray}
\eta &=& \gamma^\ast = - \frac{\epsilon}{6} \left\{ 1 + \left[ \frac{25}{288} + \frac{161}{144}\ln \left( \frac{4}{3} \right) \right] \epsilon \right\} \, ,
\\
z &=& 2 + \zeta_\rho^\ast = 2 - \frac{\epsilon}{12} \left\{ 1 + \left[ \frac{67}{288} + \frac{59}{144}\ln \left( \frac{4}{3} \right) \right] \epsilon \right\} \, ,
\\
\nu_\perp &=& \frac{1}{2-\kappa^\ast } = \frac{1}{2} + \frac{\epsilon}{16} \left\{ 1 + \left[ \frac{107}{288} - \frac{17}{144}\ln \left( \frac{4}{3} \right) \right] \epsilon \right\} \, ,
\end{eqnarray}
\end{subequations}
are the well known the critical exponents for DP which have been calculated previously to second order in $\epsilon$~\cite{janssen_81,janssen_2000}. In addition, we introduced in Eq.~(\ref{scaling}) the resistance exponent
\begin{eqnarray}
\label{resPhi}
\phi = \nu_\perp \left( 2 - \zeta_w^\ast \right) = 1 + \frac{\epsilon}{24} \left\{ 1 + \left[ \frac{151}{288} - \frac{157}{144}\ln \left( \frac{4}{3} \right) \right] \epsilon \right\} \, .
\end{eqnarray}

Equation~(\ref{scaling}) implies the following scaling behavior of the two point correlation function $G=G_2$ at criticality:
\begin{eqnarray}
\label{scaleRel}
&&G \left( |{\rm{\bf x}}_\perp-{\rm{\bf x}}_\perp^\prime |, t-t^\prime , w \vec{\lambda}^2 \right) 
\nonumber \\
&& \, = 
l^{d_\perp +\eta} G \left( l |{\rm{\bf x}}_\perp-{\rm{\bf x}}_\perp^\prime|,l^z \left( t - t^\prime \right) , l^{-\phi/\nu_\perp} w \vec{\lambda}^2 \right) \, ,
\end{eqnarray}
where we dropped several arguments for notational simplicity. In the following we set ${\rm{\bf x}}_\perp^\prime = {\rm{\bf 0}}$ and $t^\prime$, again for the sake of simplicity. The choice $l = |{\rm{\bf x}}_\perp|^{-1}$ and a Taylor expansion of the right hand side of 
Eq.~(\ref{scaleRel}) lead to 
\begin{eqnarray}
G \left( |{\rm{\bf x}}_\perp|, t , w \vec{\lambda}^2 \right) &=& \left| {\rm{\bf x}}_\perp \right|^{1-d-\eta} f_1 \left(  \frac{t}{\left| {\rm{\bf x}}_\perp \right|^z} \right) 
\nonumber \\
&\times&
\bigg\{ 1 + w \vec{\lambda}^2 \left| {\rm{\bf x}}_\perp \right|^{\phi /\nu_\perp} f_{w,1} \left(  \frac{t}{\left| {\rm{\bf x}}_\perp \right|^z} \right) + \cdots \bigg\} 
\end{eqnarray}
with $f_1$ and $f_{w,1}$ being scaling functions which vanish for vanishing argument. Equally well we can choose $l=t^{-1/z}$ which then leads to
\begin{eqnarray}
G \left( |{\rm{\bf x}}_\perp|, t , w \vec{\lambda}^2 \right)
&=& t^{(1-d-\eta )/z} f_2 \left(  \frac{\left| {\rm{\bf x}}_\perp \right|^z}{t} \right) 
\nonumber \\
&\times&
\bigg\{ 1 + w \vec{\lambda}^2 \, t^{\phi /\nu_\parallel} f_{w,2} \left(  \frac{\left| {\rm{\bf x}}_\perp \right|^z}{t} \right) + \cdots \bigg\}
\, ,
\end{eqnarray}
where $\nu_\parallel = \nu_\perp z$. $f_2$ and $f_{w,2}$ are other scaling functions which tend to constants as their argument is taken to zero. Exploiting Eq.~(\ref{shit}) we deduce that
\begin{eqnarray} 
M_R \sim t^{\phi /\nu_\parallel}
\end{eqnarray}
if measured along the preferred direction. For measurements in other directions it is appropriate to choose a length scale $L$ and to express the longitudinal and the transverse coordinates in terms of $L$: $\left| {\rm{\bf x}}_\perp \right| \sim L$ and $x_\parallel \sim L^z \sim T$. With this choice the scaling function $f_{w,1}$ reduces to a constant and we obtain 
\begin{eqnarray} 
M_R \sim L^{\phi /\nu_\perp} \sim T^{\phi /\nu_\parallel} \, .
\end{eqnarray}

\subsection{Comparison to numerical data}
\label{comparisonToData}
Now we compare our result for $\phi$ to numerical results available in the literature. We are not aware of any numerical results for $\phi$ itself. However, several authors determined the conductivity exponent $t$ governing the increase of the macroscopic conductivity $\Sigma \sim (p-p_c )^t$. The conductivity exponent is related to the resistance exponent via the scaling relation~\cite{red_83} $t = \phi + (d-1)\nu_\perp - \nu_\parallel$. Redner and Mueller~\cite{redner_mueller_82} determined $t$ in two dimensions by Monte Carlo simulations: $t (d=2) = 0.6 \pm 0.10$. Arora {\em et al}.~\cite{arora&co_83} did analogue and numerical simulations leading to $t(d=2) = 0.73 \pm 0.10$. Another value for comparison is $t(d=2) \approx 0.7$, obtained by Redner~\cite{red_83} from a real space renormalization group calculation.

Crudely evaluating the $\epsilon$-expansion in Eq.~(\ref{resPhi}) for small spatial dimensions leads inevitably to poor quantitative predictions. Therefore it is appropriate to improve the $\epsilon$-expansion by incorporating rigorously known features. We carry out a rational approximation which takes into account that $t (d=1) =0$. Practically this is done by adding an appropriate third order term. We obtain the interpolation formula
\begin{eqnarray}
\label{tAppr}
t \approx  \left( 1 - \frac{\epsilon}{4} \right) \left( 2 + 0.2083 \, \epsilon + 0.0604 \, \epsilon^2 \right) \, ,
\end{eqnarray}
which leads to $t (d=2)\approx 0.8$.

\section{Concluding remarks}
\label{concusions}
The field theory presented in this paper was tailored to describe electric transport on DP clusters. However, it is also suitable to study purely geometric aspects of DP. Our Hamiltonian $\mathcal{H}$ provides alternative means to derive the usual critical exponents of the DP universality class. For $w \to 0$ it leads to the same perturbation expansion as obtained in Refs.~\cite{janssen_81,janssen_2000}.

Our real-world interpretation of Feynman diagrams, developed in the context of RRN, proves to be a powerful tool in studying transport in DP. Here we saw that it allowed for an elegant and intuitive way of determining the average resistance $M_R$. However, the real-world interpretation is not only fruitful in studying $M_R$. It also turns out to be beneficial in investigating multifractality in DP~\cite{stenull_janssen_2001_b}
and in calculating fractal dimensions of DP clusters~\cite{janssen_stenull_directedLetter_2000,stenull_janssen_2001_c}.

The resistance exponent $\phi$ is found to be larger than the corresponding resistance exponent for the random resistor network (RRN)~\cite{lubensky_wang_85,stenull_janssen_oerding_99}. This is intuitively plausible since long tortuous paths that contribute to the macroscopic conductance in RRN are suppressed in DP.

Our result for $\phi$ is for dimensions close to five the most accurate analytic estimate that we know of. In two dimensions our result shows reasonable agreement with the known numerical results. It is certainly desirable to have more and firmer numerical data for comparison with our analytic result, in particular in three dimensions. We hope that this paper triggers further simulations of transport in DP.

\begin{acknowledgments}
We acknowledge the support by the Sonderforschungsbereich 237 ``Unordnung und gro{\ss}e Fluktuationen'' of the Deutsche Forschungsgemeinschaft. 
\end{acknowledgments}  

\appendix
\section{Calculation of the conducting diagrams}
\label{app:calculation}
This appendix outlines the calculation of the conducting Feynman diagrams listed in Fig.~\ref{fig1} in terms of examples. For briefness we focus on those parts of the diagrams proportional to $w \vec{\lambda}^2$. For the remaining parts the reader is referred to the literature on the field theory of DP~\cite{janssen_81,janssen_2000}.

We start with diagram $\mbox{A}$ which stands for
\begin{eqnarray}
\mbox{A} &=& \frac{\rho^2 g^2}{2} \int_0^\infty dt \int_{\brm{p}} \sum_{\vec{\kappa}} 
\nonumber \\
&\times& \exp \left\{ -\rho t \left[ 2\tau + 2\brm{p}^2 + w \vec{\kappa}^2 + w \big( \vec{\kappa} - \vec{\lambda} \big)^2 \right] \right\} \, ,
\end{eqnarray}
where $\int_{\brm{p}}$ is an abbreviation for $(2\pi)^{-d_\perp} \int d^{d_\perp}p$. As argued in Sec.~\ref{resistanceOfFeynmanDiagrams} the summation over the loop currents $\vec{\kappa}$ can be carried out by determining the total resistance of the diagram. Since $\mbox{A}$ is composed of two resistors of resistance $t$ in parallel its total resistance is given by 
\begin{eqnarray}
R (t) = t/2 \, .
\end{eqnarray}
Thus, we find by expanding for small $w \vec{\lambda}^2$ [cf.\ Eq.~(\ref{expansionOfDiagrams})] that $\mbox{A}$ takes the form
\begin{eqnarray}
\mbox{A} &=& - w \vec{\lambda}^2 \frac{\rho^2 g^2}{2} \int_0^\infty dt \, \rho \frac{t}{2} \exp \left( -2 \rho t \tau \right) \int_{\brm{p}}\exp \left( -2 \rho t \brm{p}^2 \right) \, .
\end{eqnarray}
Carrying out the momentum integration is straightforward. We obtain
\begin{eqnarray}
\mbox{A} &=& - w \vec{\lambda}^2 \frac{\rho g^2}{16} \frac{1}{(4\pi)^{d_\perp /2}} \int_0^\infty dt \, t^{1-d_\perp /2} \exp \left( - t \tau \right)
\nonumber \\
&=&  - w \vec{\lambda}^2 \frac{\rho g^2}{16} \frac{\Gamma (2-d_\perp /2)}{(4\pi)^{d_\perp /2}} \tau^{2-d_\perp /2} \, .
\end{eqnarray}
Expansion for small $\epsilon = 4 - d_\perp$ than finally leads to the result stated in Eq.~(\ref{resA}). Diagram $\mbox{B}$ can be calculated by the same means. One simply has to substitute $R(t)=t$ for the total resistance.

Now we turn to diagram $\mbox{H}$ which stands for 
\begin{eqnarray}
\mbox{H} &=& \frac{\rho^4 g^4}{2} \int_0^\infty dt_1 dt_2 dt_3 \int_{\brm{p},\brm{k}} \sum_{\vec{\kappa},\vec{\pi}} 
\nonumber \\
&\times& \exp \left\{ -\rho \left( t_1 + t_2 \right) \left[ 2\tau + 2\brm{p}^2 + w \vec{\kappa}^2 + w \big( \vec{\kappa} - \vec{\lambda} \big)^2 \right] \right\} 
\nonumber \\
&\times& \exp \Big\{ -\rho t_3 \Big[ 3\tau + \brm{p}^2 + \brm{k}^2 + \big( \brm{p} - \brm{k} \big)^2 
\nonumber \\
&+& w \vec{\pi}^2 + w \big( \vec{\kappa} - \vec{\pi} \big)^2 + w \big( \vec{\kappa} - \vec{\lambda} \big)^2 \Big] \Big\}
\, .
\end{eqnarray}
By applying the usual rules for determining the total resistance of resistors added in parallel or in series we find that the total resistance of $\mbox{H}$ is given by
\begin{eqnarray}
R \left( t_1 ,t_2 ,t_3 \right) = \frac{t_1^2 + t_2^2 + \frac{1}{2} t_3^2 + 2 t_1 t_2 + \frac{3}{2} t_1 t_3 + \frac{3}{2} t_2 t_3}{2t_1 + 2t_2 + \frac{3}{2}t_3}
\, .
\end{eqnarray}
The momentum integrations are easily carried out by completing squares. After these steps we arrive at
\begin{eqnarray}
\mbox{H} &=& -w \vec{\lambda}^2 \frac{\rho g^4}{2} \frac{1}{(4\pi)^{d_\perp}} \int_0^\infty dt_1 dt_2 dt_3 \exp \left[ -\tau \left( 2t_1 +2t_2 +3t_3 \right) \right]
\nonumber \\
&\times& \frac{1}{(2t_3)^{d_\perp /2}} \, \frac{t_1^2 + t_2^2 + \frac{1}{2} t_3^2 + 2 t_1 t_2 + \frac{3}{2} t_1 t_3 + \frac{3}{2} t_2 t_3}{\left[ 2t_1 + 2t_2 + \frac{3}{2}t_3 \right]^{1+d_\perp /2}} \, .
\end{eqnarray}
The remaining integrations can be simplified by changing variables: $t_1 \to \frac{1}{2} ty$, $t_2 \to \frac{1}{2} t (1-x-y)$ and $t_3 \to \frac{1}{3} tx$. This gives
\begin{eqnarray}
\mbox{H} &=& -w \vec{\lambda}^2 \, \frac{\rho g^4}{12} \, \frac{\Gamma (4-d_\perp)}{(4\pi)^{d_\perp}} \tau^{d_\perp -4} \left( \frac{3}{4} \right)^{d_\perp /2} \int_0^1 dx \int_0^{1-x} dy   
\nonumber \\
&\times& 
 x^{-d_\perp /2} \left[ 1 + x \right]^{d_\perp -4}
\bigg\{ \frac{1}{4} y^2 + \frac{1}{4} (1-x-y)^2 + \frac{2}{9} x^2 
\nonumber \\
&+& \frac{1}{2} y(1-x-y) + \frac{1}{2} xy + \frac{1}{2} x(1-x-y) \bigg\} \, .
\end{eqnarray}
where we have already carried out the integration over $t$. The integration over $y$ can be looked up in a table~\cite{gradshteyn_ryzhik}. We obtain
\begin{eqnarray}
\label{SourceOfIs}
\mbox{H} &=& -w \vec{\lambda}^2 \, \frac{\rho g^4}{12} \, \frac{\Gamma (4-d_\perp)}{(4\pi)^{d_\perp}} \tau^{d_\perp -4} \left( \frac{3}{4} \right)^{d_\perp /2}    
\nonumber \\
&\times& \bigg\{ \frac{2}{9} I_1 +  \frac{1}{2} I_2 + \frac{1}{4} I_3 \bigg\} \, ,
\end{eqnarray}
where
\begin{eqnarray}
I_1 &=& \int_0^1 dx \, x^{2-d_\perp /2} \left[ 1 + x \right]^{d_\perp -4} (1-x) \, ,
\\
I_2 &=& \int_0^1 dx \, x^{1-d_\perp /2} \left[ 1 + x \right]^{d_\perp -4} (1-x)^2 \, ,
\\
I_3 &=& \int_0^1 dx \, x^{-d_\perp /2} \left[ 1 + x \right]^{d_\perp -4} (1-x)^3
\, .
\end{eqnarray}
The integral $I_1$ is convergent for $d_\perp \leq 4$. Hence it is sufficient to evaluate it at the upper critical dimension, i.e., for  $d_\perp = 4$. Then we immediately obtain that
\begin{eqnarray}
\label{I1}
I_1 = 1/2
\, .
\end{eqnarray}
Integral $I_2$ can be evaluated by separating its divergent and convergent contributions:
\begin{eqnarray}
\label{I2}
I_2 &=& \int_0^1 dx \left\{ x^{1-d_\perp /2} + x^{-1} \left[ (1 - x)^2 -1 \right]^2 \right\} 
\nonumber \\
&=& \frac{2}{4-d_\perp} - \frac{3}{2}
\, .
\end{eqnarray}
$I_3$ may be treated by the same means as $I_2$ giving
\begin{eqnarray}
\label{I3}
I_3 =  \frac{2}{2-d_\perp} + \frac{2(d_\perp -7)}{4-d_\perp} + \frac{5}{2}
\, .
\end{eqnarray}
After substituting Eqs.~(\ref{I1}) to (\ref{I3}) into Eq.~(\ref{SourceOfIs}) we carry out an expansion for small $\epsilon$ which leads to the result stated in Eq.~(\ref{resH}). Diagrams $\mbox{I}$ to  $\mbox{L}$ can be evaluated in an analogous fashion.

As a third and final example we consider diagram $\mbox{C}$ standing for
\begin{eqnarray}
\mbox{C} &=& \rho^4 g^4 \int_0^\infty dt_1 dt_2 dt_3 \int_{\brm{p},\brm{k}} \sum_{\vec{\kappa},\vec{\pi}}
\nonumber \\
&\times& \exp \left\{ -\rho t_1 \left[ 2\tau + 2\brm{p}^2 + w \vec{\kappa}^2 + w \big( \vec{\kappa} - \vec{\lambda} \big)^2 \right] \right\}
\nonumber \\
&\times& \exp \left\{ -\rho t_2 \left[ 2\tau + 2\brm{k}^2 + w \vec{\pi}^2 + w \big( \vec{\pi} - \vec{\lambda} \big)^2 \right] \right\}
\nonumber \\
&\times& \exp \Big\{ -\rho t_3 \Big[ 3\tau + \brm{p}^2 + \brm{k}^2 + \big( \brm{p} - \brm{k} \big)^2 
\nonumber \\
&+& w \vec{\pi}^2 + w \big( \vec{\kappa} - \vec{\pi} \big)^2 + w \big( \vec{\kappa} - \vec{\lambda} \big)^2 \Big] \Big\}\nonumber \\ 
\, .
\end{eqnarray}
Solving Kirchhoff's equations for this diagram results in
\begin{eqnarray}
R \left( t_1 ,t_2 ,t_3 \right) &=& \frac{1}{\left[4 (t_1 + t_3)(t_2 + t_3) - t_3^2\right]^2}
\nonumber \\
&\times&
\Big\{
t_1 \left[  2 t_1 t_2 + 2 t_1 t_3 + 3 t_2 t_3 + 2 t_3^2 \right]^2
\nonumber \\
&+&
t_2 \left[  2 t_1 t_2 + 2 t_2 t_3 + 3 t_1 t_3 + 2 t_3^2 \right]^2
\nonumber \\
&+&
\left( t_1 + t_3 \right) \left[  2 t_1 t_2 + 2 t_1 t_3 + t_2 t_3 +  t_3^2 \right]^2
\nonumber \\
&+&
\left( t_2 + t_3 \right) \left[  2 t_1 t_2 + 2 t_2 t_3 + t_1 t_3 +  t_3^2 \right]^2
\nonumber \\
&+&
t_3 \left[ t_1 t_3 + t_2 t_3 + t_3^2 \right]^2
\Big\}
\, .
\end{eqnarray}
Hence the summations over the loop currents lead in conjunction with the momentum integrations to 
\begin{eqnarray}
\mbox{C} = - w \vec{\lambda}^2 \rho g^4 \big\{ M_1 + 2 M_2 + 2 M_3 \big\} 
\, ,
\end{eqnarray}
where
\begin{eqnarray}
M_1 &=& \frac{1}{(4\pi)^{d_\perp}} \int_0^\infty dt_1 dt_2 dt_3 \, \exp \left[ -\tau \left( 2t_1 +2t_2 +3t_3 \right) \right]
\nonumber \\
&\times& t_3^3 \, \frac{\left[ t_1 + t_2 + t_3 \right]^2}{\left[4 (t_1 + t_3)(t_2 + t_3) - t_3^2\right]^2}
\, ,
\\
M_2 &=& \frac{1}{(4\pi)^{d_\perp}} \int_0^\infty dt_1 dt_2 dt_3 \, \exp \left[ -\tau \left( 2t_1 +2t_2 +3t_3 \right) \right]
\nonumber \\
&\times& t_1 \, \frac{\left[ 2 t_1 t_2 + 2 t_1 t_3 + 3 t_2 t_3 + 2 t_3^2\right]^2}{\left[4 (t_1 + t_3)(t_2 + t_3) - t_3^2 \right]^2}
\, ,
\\
M_3 &=& \frac{1}{(4\pi)^{d_\perp}} \int_0^\infty dt_1 dt_2 dt_3 \, \exp \left[ -\tau \left( 2t_1 +2t_2 +3t_3 \right) \right]
\nonumber \\
&\times& \left( t_1 + t_3 \right) \frac{\left[ 2 t_1 t_2 + 2 t_1 t_3 + t_2 t_3 +  t_3^2 \right]^2}{\left[4 (t_1 + t_3)(t_2 + t_3) - t_3^2 \right]^2}
\, .
\end{eqnarray}
The integral $M_1$ can be simplified by the following change of variables: $t_1 \to t(x-1)$, $t_2 \to t(y-1)$, and $t_3 \to t$. Upon carrying out the integration over $t$ we then obtain
\begin{eqnarray}
M_1 &=& \frac{\Gamma ( 4-d_\perp )}{(4\pi)^{d_\perp}} \tau^{d_\perp -4} \int_1^\infty dx dy 
\nonumber \\
&\times& \frac{\left[ 2x + 2y  -1 \right]^{d_\perp -4} \left[ x + y -1 \right]^2}{\left[ 4 xy -1 \right]^{2 + d_\perp /2}}
\nonumber \\
&=&
\frac{\Gamma ( 4-d_\perp )}{(4\pi)^{d_\perp}} \tau^{d_\perp -4} 2^{-7} \int_1^\infty dx \int_1^x dy
 \nonumber \\
&\times& \frac{\left[ x + y  - \frac{1}{2} \right]^{d_\perp -4} \left[ x + y -1 \right]^2}{\left[ xy - \frac{1}{4} \right]^{2 + d_\perp /2}}
\, .
\end{eqnarray} 
Since the remaining integrations are convergent for $d_\perp \leq 4$ we evaluate them directly at $d_\perp = 4$ which gives in $\epsilon$-expansion
\begin{eqnarray}
M_1 = - \frac{G_\epsilon^2}{1728 \epsilon} \, \tau^{-\epsilon} \left\{ 25 -108 \ln \left( \frac{4}{3} \right)  \right\}
\, .
\end{eqnarray} 
$M_2$ and $M_3$ can be analyzed similarly. However, these calculations are somewhat more tedious. After all one obtains for $\mbox{C}$ the result stated in Eq.~(\ref{resC}). Diagrams $\mbox{D}$ to $\mbox{G}$ may be computed like $\mbox{C}$.

\section{Results for the conducting diagrams}
\label{app:results}
Here we list our results for the conducting diagrams. For briefness we display only those parts of the diagrams proportional to $w \vec{\lambda}^2$: 
\begin{eqnarray}
\label{resA}
\mbox{A} &=& - w \vec{\lambda}^2 \rho g^2 \, \frac{G_\epsilon}{8 \, \epsilon} \, \tau^{-\epsilon /2} \ ,
\\
\mbox{B} &=& - w \vec{\lambda}^2 \rho g^2 \, \frac{G_\epsilon}{4 \, \epsilon} \, \tau^{-\epsilon /2} \ ,
\\
\label{resC}
\mbox{C} &=& - w \vec{\lambda}^2 \rho g^4 \, \frac{G_\epsilon^2}{8 \, \epsilon} \, \tau^{-\epsilon} \left[ \frac{1}{\epsilon} + \frac{7}{12} - \ln \left( \frac{4}{3} \right) \right] \ ,
\\
\mbox{D} &=& - w \vec{\lambda}^2 \rho g^4 \, \frac{G_\epsilon^2}{16 \, \epsilon} \, \tau^{-\epsilon} \left[ \frac{3}{\epsilon} + \frac{3}{2} - \frac{5}{2} \ln \left( \frac{4}{3} \right) \right] \ ,
\\
\mbox{E} &=& - w \vec{\lambda}^2 \rho g^4 \, \frac{G_\epsilon^2}{8 \, \epsilon} \, \tau^{-\epsilon} \left[ \frac{1}{\epsilon} + \frac{1}{2} - \frac{1}{2} \ln \left( \frac{4}{3} \right) \right] \ ,
\\
\mbox{F} = \mbox{G} &=& - w \vec{\lambda}^2 \rho g^4 \, \frac{G_\epsilon^2}{8 \, \epsilon} \, \tau^{-\epsilon} \left[ \frac{2}{\epsilon} + 1 -  \ln \left( \frac{4}{3} \right) \right] \ ,
\\
\label{resH}
\mbox{H} &=&  w \vec{\lambda}^2 \rho g^4 \, \frac{3 \, G_\epsilon^2}{128 \, \epsilon} \, \tau^{-\epsilon} \left[ \frac{1}{\epsilon} + \frac{55}{36} + \frac{1}{2} \ln \left( \frac{4}{3} \right) \right] \ ,
\\
\mbox{I} &=&  w \vec{\lambda}^2 \rho g^4 \, \frac{G_\epsilon^2}{16 \, \epsilon} \, \tau^{-\epsilon} \left[ \frac{1}{\epsilon} + 1 + \frac{1}{2} \ln \left( \frac{4}{3} \right) \right] \ ,
\\
\mbox{J} &=&  w \vec{\lambda}^2 \rho g^4 \, \frac{G_\epsilon^2}{128 \, \epsilon} \, \tau^{-\epsilon} \left[ \frac{2}{\epsilon} + 5 + \ln \left( \frac{4}{3} \right) \right] \ ,
\\
\mbox{K} = \mbox{L} &=&  w \vec{\lambda}^2 \rho g^4 \, \frac{G_\epsilon^2}{64 \, \epsilon} \, \tau^{-\epsilon} \left[ \frac{2}{\epsilon} + 5 + \ln \left( \frac{4}{3} \right) \right] \ .
\end{eqnarray}


\end{document}